# Impact of parallel code optimization on computer power consumption

E. A. Kiselev,[1, *] P. N. Telegin,[1, **] and A. V. Baranov[1, ***]

[1]*Joint Supercomputer Center of the Russian Academy of Sciences – Branch of Federal State Institution "Scientific Research Institute for System Analysis of the Russian Academy of Sciences", Leninsky prospect, 32a, Moscow, 119334, Russian Federation*

Abstract— The increase in performance and power of computing systems requires the wider use of program optimizations. The goal of performing optimizations is not only to reduce program runtime, but also to reduce other computer resources including power consumption. The goal of the study was to evaluate the impact of different optimization levels and various optimization strategies on power consumption. In a series of experiments, it was established that the average power consumption tends to peak for the programs with optimized source code. The articles also describes the impact of changing computer architecture on power consumption graphs. The relationships between the average and median values of power consumption by example programs are considered. The possibility of creating program energy consumption profile for a parallel program is shown.

* E-mail: kiselev@jscc.ru

** E-mail: ptelegin@jscc.ru

*** E-mail: abaranov@jscc.ru

## 1. INTRODUCTION

The range of tasks and the scope of application of computing clusters expands every year. Due to high performance and scalability of clusters, it is possible to reduce significantly the time of experimental research and increase the accuracy of modeling. However, the use of more efficient computing systems does not always lead to reduction in the parallel programs execution time. This effect is often associated with the low parallelism of the algorithm, the incompatibility of the parallel algorithm with the computer architecture, or the presence of sections of program code that cannot be vectorized. Typically, it is needed to rework the parallel algorithm partially or completely to eliminate the first two causes. This requires considerable time and is not always possible. Problems with program code vectorization can happen due to dependencies that can be resolved without changing the program algorithm [1]. For this kind of tasks the use of various program code optimization techniques allows to reduce the execution time of a parallel application, as well as the computer power consumption.

Any optimization has a side effect in increased load of the CPU, GPU, memory and disk subsystem. The high combined power consumption of all these components during peak periods leads to slowing down performance or even to faults and failures in the computer operation. This problem is typical for cluster computing systems, when computing nodes are operating in the energy «red zones». That is why the evaluation of various strategies of program code optimization for peak power consumption is relevant.

## 2. PARALLEL PROGRAM POWER CONSUMPTION PROFILE

Depending on the parallel programming model, the program can run on one or several supercomputer nodes. Regardless of the implemented algorithm, the following phases of parallel program execution can be identified: computation, disk subsystem access, communication exchanges between computer nodes. The execution time of each parallel program phase and the amount of consumed energy depends on the characteristics of the computing system.

Let us denote:

$T$ is the parallel program execution time;

$N$ is the number of computing nodes allocated to the job;

$E_{calc,i}^{j}(t)$ is the amount of consumed energy by device $i$ of the node $j$ at time point $t$, where

$$t \in [1;T], j \in [1;N];$$

$E_{CALC,\Sigma}^{j}(t)$ is the amount of energy consumed by all computing devices of node $j$ at time point $t$:

$$E_{CALC,\Sigma}^{j}(t) = \sum_{i} E_{calc,i}^{j}(t),\ i = \{CPU, GPU, \ldots\}$$

$E_{disk}^{j}(t)$ is the amount of energy consumed by node $j$ when accessing the external memory at time point $t$;

$E_{net}^{j}(t)$ is the of energy consumed by node $j$ when sending/receiving data through the communication network at time point $t$;

$W^{j}(t)$ is the power consumption of node $j$ at moment $t$:

$$W_j(t) = E_{CALC,\Sigma}^{j}(t) + E_{disk}^{j}(t) + E_{net}^{j}(t)$$

Maximum value of power consumption of node $j$ for running parallel program (J/s) $W_{max,j}$, can be found using the formula:

$$W_{max,j} = max_{t \in [1,T]} W_j(t),\ j \in [1,N]$$

When scheduling the start of an application and determining the list of allocated computing resources, it is important to control its impact on the energy consumption of the set of allocated computers. The worst case scenario is the case when, the maximum energy consumption value $W_{max,j}$ is reached on all allocated nodes, and the total power consumption is:

$$W_{max} = \Sigma_j W_{max,j},\ j \in [1,N]$$

The feasibility of using peak power consumption values $W_{max}$ derives from the increased probability of exceeding power consumption threshold and the emergency shutdown of the computer at such values.

Besides monitoring single peak values, it is important to evaluate the load on the power grid of a computer system during the parallel program runtime. Previous studies show that the median value of power consumption is best suited for such evaluation. This value we denote as $W_{med}$ (J/S), and calculate it as follows.

Let us measure the power consumption n times at moments $1, 2, \ldots, n$. We denote instantaneous value of power consumption at time $i$ as $x_i$. Let us sort number series $x_1, \ldots, x_n$ in ascending order. We get a new series $x_1^{'}, \ldots, x_n^{'}$. The median power consumption is calculated as follows:

$$E_{med} = x'_{(n+1)/2} \text{ for odd } n, \quad E_{med} = \frac{x'_{n/2} + x'_{(n+1)/2}}{2} \text{ for even } n$$

The pair $\{W_{med}, W_{max}\}$, describes the influence of an application on the computer power consumption. Let us refer this pair as a power consumption profile. We measure this profile as the median and maximum energy spent by the computer during the execution of the application.

## 3. OPTIMIZATION METHODS FOR PARALLEL PROGRAM CODE

Modern compilers support various program optimization strategies. For example, different versions of gcc compiler [2, 3] support optimization flags -O0, -O1 (-O), -O2, -O3, each of which contains certain set of program optimization methods. By default or with -O0 flag the compiler reduces compilation time and make debugging produce the expected results. Flags -O1 (-O), -O2, -O3 makes the compiler to use different program optimizations to reduce program execution time taking more time, and more memory.

The most used optimization methods are the following.

1. Function inlining
2. Constant folding and constant propagation
3. Pointer elimination
4. Common subexpression elimination
5. Register variables
6. Loop invariant code motion
7. Anti- and Output dependences elimination in loop
8. Optimization of memory allocation inside loop

Let us consider these optimizations.

### *3.1. Function inlining*

The idea of this optimization way is replacing a function call with the body of the called function. This reduces overheads for function call and return, and arguments passing. Cache is used more efficiently, and due to linear code other optimizations can be performed more efficiently. An example of this method is presented in the Listing 1.

Listing 1.

```
// Before optimization
float square (float a) {
  return a*a;
}
float parabola (float x) {
  return square(x) + 1.0f;
}
```

```
// After optimization
float parabola (float x) {
   return x*x + 1.0f;
}
```

In this example the function body is placed next to call site. When the function is called from another module the compiler creates a copy of the functions. This results in fragmentation of program code and reduces efficiency of instruction cache. To exclude this kind of cases one should declare function as static, and use compiler options, which make linker to exclude all functions without calls.

### *3.2. Constant folding and constant propagation*

This optimization substitutes constant expression with its value. In the Listing 2 there are three examples of compiler optimization.

Listing 2.

```
// Example 1:
a = b + 2.0/3.0; // Before optimization        a = b + 0.666666666666667; // After optimization
// Example 2:
b*2.0/3.0; // Before optimization              (b*2.0)/3.0; // After optimization
// Example 3:
//Before optimization                          //After optimization
a = (2.0/3.0);                                 a = 0.666666666666667;
b = a + 1;                                     b = 1.666666666666667;
```

If function call cannot be inlined or calculated at compilation time, then constant folding and propagation cannot be performed. To change this situation, it is needed to substitute function call with function body (see Listing 3).

Listing 3.

```
// Before optimization                 // After optimization
float parabola (float x) {             float a, b;
  return x*x + 1.0f;                   a = 5.0f;
}                                      b = 6.0f;
float a, b;
a = parabola (2.0f);
b = a + 1.0f;
```

### *3.3. Pointer elimination*

This optimization reduces number of pointers when possible (see Listing 4).

Listing 4.

```
// Before optimization void            // After optimization
Plus2 (int *p) {                       a += 2;
  *p = *p + 2;
}
int a;
Plus2 (&a);
```

### *3.4. Common subexpression elimination*

If there is more than one occurrence of the identical expression, and it is possible to substitute with a variable, then the compiler can compute it once and use whenever it is reasonable (see Listing 5).

Listing 5.

```
// Before optimization                 // After optimization
int a, b, c;                           int a, b, c, temp;
b = (a+1) * (a+1);                     temp = a+1;
c = (a+1) / 4;                         b = temp * temp;
                                       c = temp / 4;
```

### *3.5. Register variables*

Compiler allocates the most often uses variables in registers (for all variable lifetime or part of it). Typically, the following data is placed in registers: intermediate values in expressions, loop counters, function arguments, pointers (including *this*), common subexpressions, induction variables.

Variable cannot be placed in register if it was declared as volatile, which means that its value can be changed outside function.

Optimizing compiler can use the same register for several variables if their ranges of use do not intersect or their values are the same.

### *3.6. Loop invariant code motion*

It operation inside loop does not depend on counter, it can be moved outside the body of a loop to reduce overhead (see Listing 6).

Listing 6.

```
// Before optimization
int i, a[100], b;
for (i = 0; i < 100; i++)
  a[i] = b * b + 1;
```

```
// After optimization
int i, a[100], b, temp;
temp = b * b + 1;
for (i = 0; i < 100; i++) {
  a[i] = temp;
}
```

### *3.7. Anti- and Output dependences elimination in loop*

If two interdependent operations are executed inside loop, they will be executed consecutively. For their parallel execution it is necessary to make variables independent on each loop iteration. Example of dependence elimination is presented in the Listing 7.

Listing 7.

```
// Before optimization
for (i = 0; i < N; i++) {
  x = a[i] + 1;
  b[i] = x*x;
}
```

```
// After optimization (option 1)
for (i = 0; i < N; i++) {
  x[i] = a[i] + 1;
  b[i] = x[i]*x[i];
}
// After optimization (option 2)
for (i = 0; i < N; i++) {
  int x = a[i] + 1;
  b[i] = x*x;
}
```

If input data for previous iterations are rewritten in a loop, then this loop cannot be vectorized. If you declare temporary variable inside loop and assign to it the value of next iteration, then loop vectorization becomes possible. In the Listing 8 one can see an example of such elimination.

Listing 8.

```
// Before optimization
for (i = 0; i < N; i++){ a[i] = b[i] + 1; c[i] = (a[i] +
a[i+1])/2; }
```

```
// After optimization (option 1)
for (i = 0; i < N; i++){
  temp[i] = a[i+1];
```

```
  a[i] = b[i] + 1;
  c[i] = (a[i] + temp[i])/2;
}
// After optimization (option 2)
for (i = 0; i < N; i++){
  float temp = a[i+1];
  a[i] = b[i] + 1;
  c[i] = (a[i] + temp)/2;
}
```

### *3.8. Optimization of memory allocation inside loop*

Access to data placed close to each other in memory is faster. We can illustrate this principle with programs presented in the Listing 9. Location in the array pointer is used.

In non-optimized version 500 array elements are read in the loop. In optimized version also 500 elements are read, but they are located in blocks.

Listing 9.

```
// Before optimization
for (int i = 0; i < 10000000; i++) {
  int sum = 0;
  for (int x = 0; x < 10000; x += 100)
    sum += values[x];
}
```

```
// After optimization
for (int i = 0; i < 10000000; i++) {
  int sum = 0;
  for (int x = 0; x < 100; x++)
    sum += values[x];
}
```

A predicate-driven group of methods for program parallelization, substitutions and code eliminations is discussed in [1].

## 4. ANALYSIS OF PREVIOUS WORKS

The problem of computer power consumption and the search for ways to reduce it is highly relevant now. One of the most common and least expensive ways to reduce the power consumption is the development of energy-efficient applications. Software developers often face the need to optimize existing program code. An analysis of works describing use of various program code optimization strategies demonstrates the feasibility of their use and the possibility of significant reduction of power consumption while reducing application execution time.

Although most of the works are focused on program code optimization targets mobile platforms, the same approaches work well on high-performance computing systems. The practical application of the methods for optimizing loops in the Java language proposed in [4, 5] demonstrates a significant reduction in power consumption. Table 1 shows our results of power consumption estimation for the optimized and non-optimized C++ code in the Listings 10–13. You can notice that although the execution time and power consumption are reduced, the peak power consumption value for the optimized code is increased.

Listing 10.

```
// Before optimization
const int n = 1000000;
for (int q = 0; q < 50000; q++) {
  int B[n];
```

```
// After optimization
const int n = 1000000;
for (int q = 0; q < 50000; q++) {
  int B[n];
```

```
  for (int i = 0; i < n; i++)
    B[i] = A[i];
}
```

```
  std::copy_n(A, n, B);
}
```

Listing 11.

```
// Before optimization
for(int i = 0; i < 50000000; i++) {
  int sum = 0;
  for (int x = 0; x < 50000; x += 100)
    sum += values[x];
}
```

```
// After optimization
for(int i = 0; i < 50000000; i++) {
  int sum = 0;
  for (int x = 0; x < 500; x++)
    sum += values[x];
}
```

Table 1. Comparison of energy consumption for C++ loops

| Listing no. | Before optimization | | | After optimization | | |
|---|---|---|---|---|---|---|
| | $W_{max}$ (J/sec) | Runtime (sec) | Power cons. (J) | $W_{max}$ (J/sec) | Runtime (sec) | Power cons. (J) |
| 10 | 13.9 | 329.7 | 4570.8 | 15.8 | 46.0 | 725.9 |
| 11 | 13.2 | 177.3 | 2327.2 | 13.6 | 158.0 | 2146.7 |
| 12 | 17.8 | 264.3 | 4714.3 | 20.1 | 4.0 | 80.0 |
| 13 | 17.4 | 310.0 | 5391.3 | 19.0 | 4.0 | 76.0 |

An example of reducing power consumption by converting three nested loops into a single one is provided in [6]. Table 1, row 3 shows our results of power consumption estimation for the optimized and non-optimized C++ code (see Listing 12). Similar to the first example, execution time and power consumption are reduced, the peak power consumption value for the optimized code is increased.

Article [7] describes a study of various strategies influence on power consumption. The authors considered the following methods for program code optimization:

R1 — using private variables instead of public ones;

R2 — using variable instead of reference;

R3 — reducing expression by addition of new variable;

R4 — transformation of loop nest into a single loop;

R5 — using delegation instead of inheritance [8].

In this case, R4 and R3 have the greatest impact on energy consumption. Table 1, row 4 shows the result of estimation of the impact on energy consumption using the (R2–R3–R4) strategy. An example of optimized and non-optimized C++ code is shown in the Listing 13.

Listing 12.

```
// Before optimization
int a = 10000, b = 10000, c = 1000, result = 0;
```

```
// After optimization
int a = 10000, b = 10000, c = 1000, result = 0;
```

```
for (int q1 = 0; q1 < a; q1++)
  for (int q2 = 0; q2 < b; q2++)
    for (int q3 = 0; q3 < c; q3++)
      result += a*b;
```

```
for (int q1 = 0; q1 < a*b*c; q1++)
  result += a*b;
```

Listing 13.

```
// Before optimization
int a = 10000, b = 10000, c = 1000, result = 0;
int &ref = a;
for (int q1 = 0; q1 < a; q1++)
  for (int q2 = 0; q2 < b; q2++)
   for (int q3 = 0; q3 < c; q3++) {
     if (ref) result += a*b;
       if (q1 < 1-100 && a+2 < q1 && b*10 != -
1)
         result += a*b;
   }
```

```
// After optimization
int a = 10000, b = 10000, c = 1000, result = 0;
for (int q1 = 0; q1 < a*b*c; q1++) {
  if (a) result += a*b;
   bool var = 1 < 1-100 && a+2 < q1 && b*10
!= -1;
   if (var) result += a*b;
}
```

Note for all examples, that the use of various strategies for the program code optimization resulted in peak energy consumption increase values by an average of 10–15% against the backdrop of a general reduction in program execution time. An increase in peak power consumption of up to 17% was established for some examples.

The article [6] shows a direct relationship of the computer energy consumption on the number of instructions executed by the microprocessor core (*Executed Instruction Count*) and the number of memory access operations (*Memory Access Count*). There is a significant impact of memory access operations during loop execution on the energy consumption. In some cases, it may exceed the energy consumption for executing core instructions. These factors indicate that parallel applications with loops processing amounts of data significantly larger than the microprocessor cache and requiring core operations, can contribute to peak power consumption values. Based on research [1, 4–7] and the results of our analysis, we can make a conclusion concerning the impact of program code optimization on peak power consumption values and the need to control them.

## 5. RESEARCH RESULTS

We conducted a series of experiments to study the influence of various program code optimization strategies on the power consumption of computers. We measured the peak and median power consumption of various computing systems for parallel applications. We tested MPI and OpenMP programs implementing parallel Dijkstra algorithms (dijkstra), parallel sort of an array of strings (sortstr), calculation of the differential equation integral by the Euler method (koshi) and Riemann differential equation (riemann). The computational complexity of each algorithm is shown in Table 2.

Each program was compiled with the optimization flags –O, –O2 and –O3 and was run on the computing resources of four JSCC RAS supercomputers: MVS–10P MP2 KNL (KNL), MVS–10P OP Broadwell (Broadwell), MVS–10P OP Skylake (Skylake) and MVS–10P OP Cascade lake (clk)[9].

We developed a module for automatic collection (*collector*) and processing of data (*exporter*) on power consumption of parallel programs, integrated into the SUPPZ workload manager [10]. The *collector* monitor program collects data on the current power consumption of the computing unit every second and saves it in a separate file for each computing unit. The powercap [11], perf_event_open [12], or MSR register [13] interfaces can be the sources for collecting power consumption data. The *exporter* program generates a single report based on the collected files, which contain statistical information on power consumption of parallel application, including the $W_{max}$ and $W_{med}$ values.

At the first stage, we estimated how the nature of energy consumption changes depending on flag –O, –O2 or –O3. To do this, we summarized the results of program runs on different computers and plotted graphs of changes in power consumption during the execution of each program (see Fig. 1–4).

In the next stage, we examined the impact of optimization on peak and median energy consumption. It was found that the difference between peak and median power consumption depends not only on the type of computing platform, but also on the optimization flag used at the compilation stage.

Changing the optimization level from –O to –O3 typically reduces the difference between median and peak power consumption (Figures 5–8). For the KNL platform, an increase in peak and median power consumption was observed when we changed the optimization flags from –O to –O3 using the MPI library. In all other cases, changing the optimization flags did not result in an increase in peak and median power consumption values, but on the contrary most often led to their decrease. The greatest differences between peak and median power consumption (from 6% to 25%) were observed for applications using the MPI library or a combination of MPI and OpenMP. When parallel applications used only the OpenMP library, the difference did not exceed 6%.

Table 2. Computational complexity of parallel algorithms

| Name of application (used libraries) | The computational complexity of algorithm |
|---|---|
| Sortstr (MPI) | $O(\frac{n}{p} log_2 \frac{n}{p} + 2n)$ ,<br>where $p$ — number of computing processes,<br>$n$ — number of array elements |
| Dijkstra (MPI + OpenMP) | $O(\frac{n^2}{p})$ ,<br>where $p$ — number of computing processes,<br>$n$ — number of graph vertices |
| Koshi (MPI + OpenMP)<br>Riemann (OpenMP) | $O\left(\frac{n^2\tau}{p}\right)$,<br>where $n > p$ , $p$ — number of computing processes<br>$n$ — number of unknowns in equation $x_1, x_2, \cdots, x_n$,<br>$\tau$ — coefficient that determines accuracy of solution and influences number of iterations |

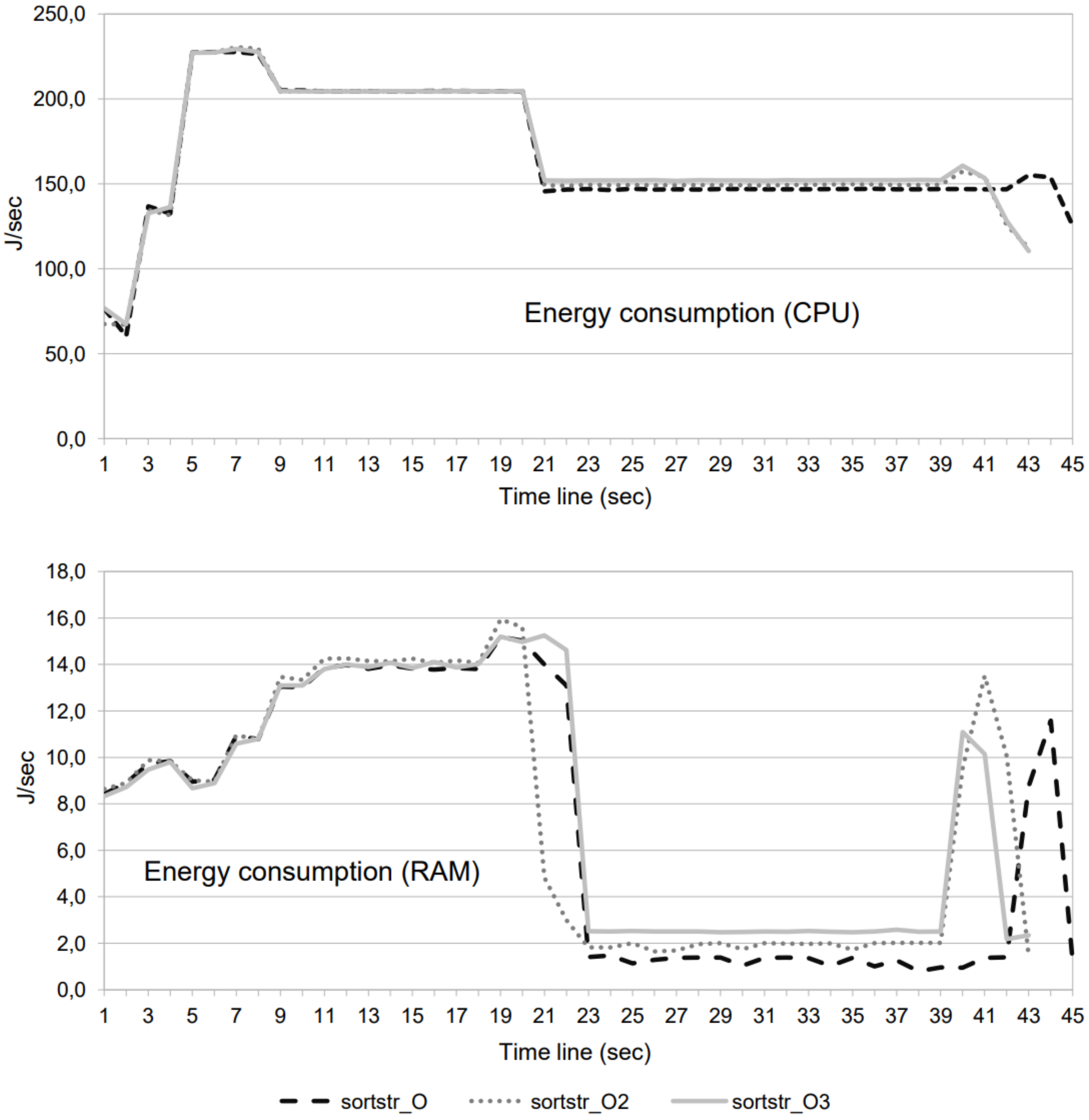

Figure 1. Profile for sortstr program power consumption changes with different optimization flags

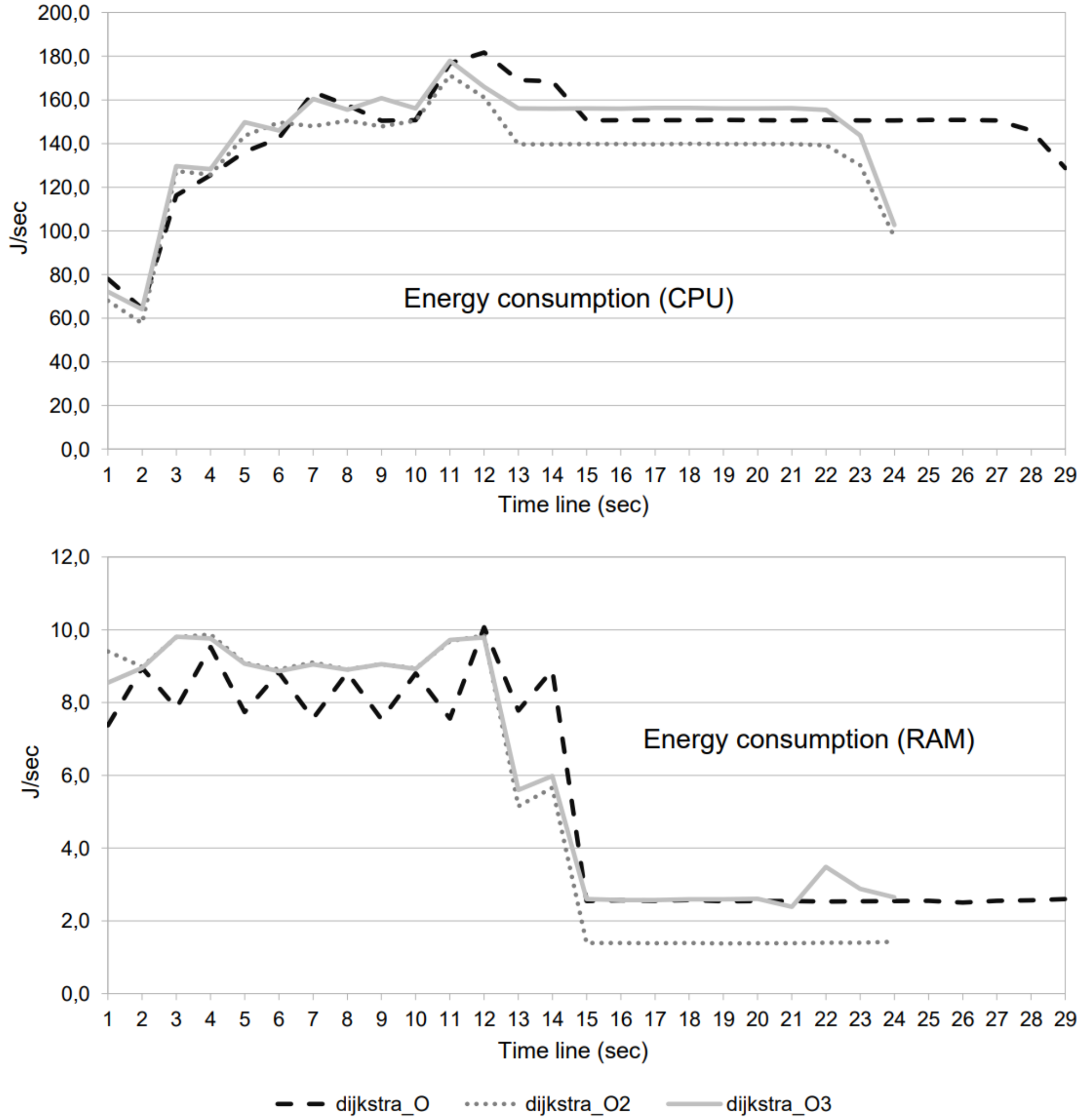

Figure 2. Profile for dijkstra program power consumption changes with different optimization flags

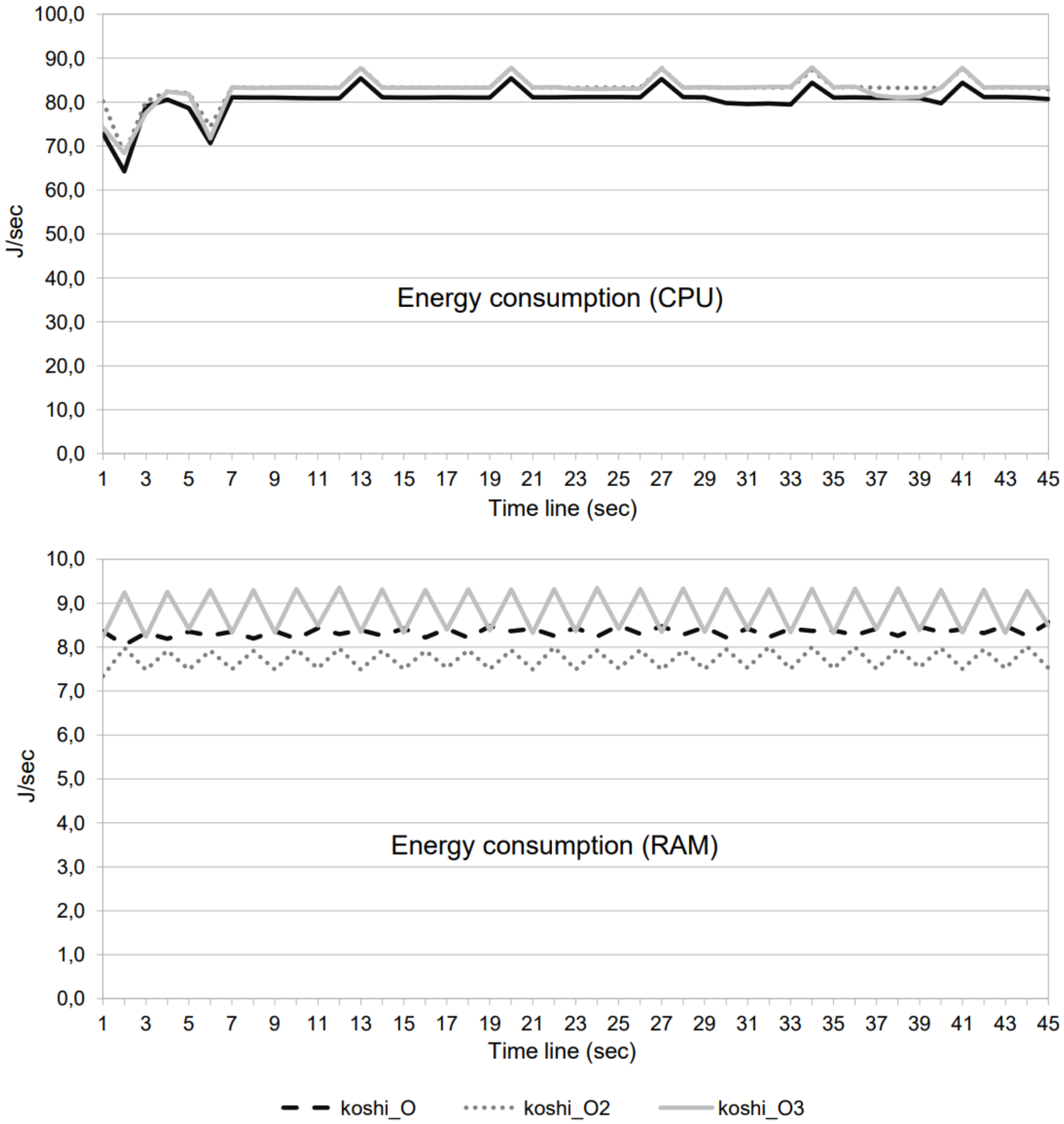

Figure 3. Profile for koshi program power consumption changes with different optimization flags

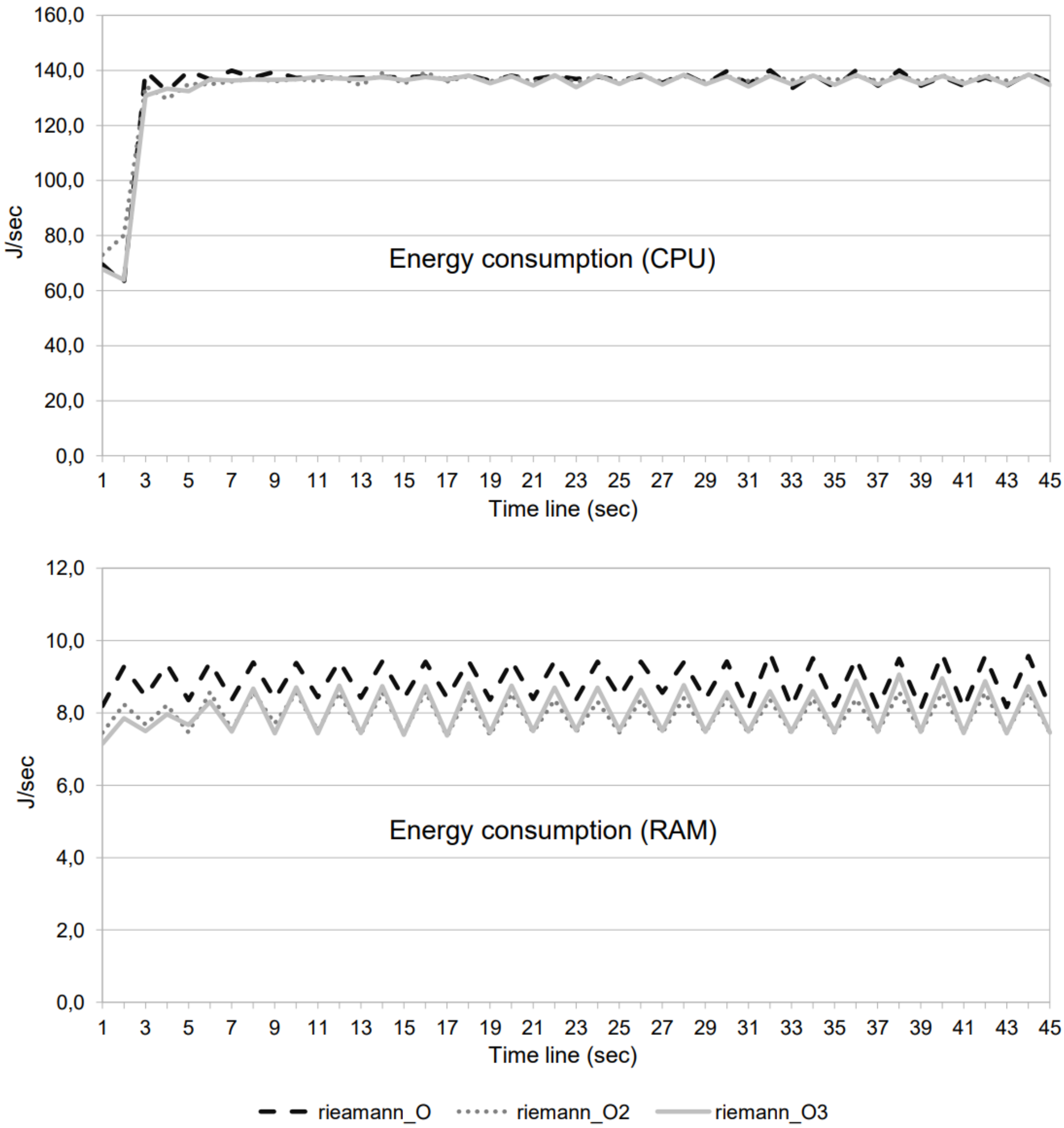

Figure 4. Profile for riemann program power consumption changes with different optimization flags

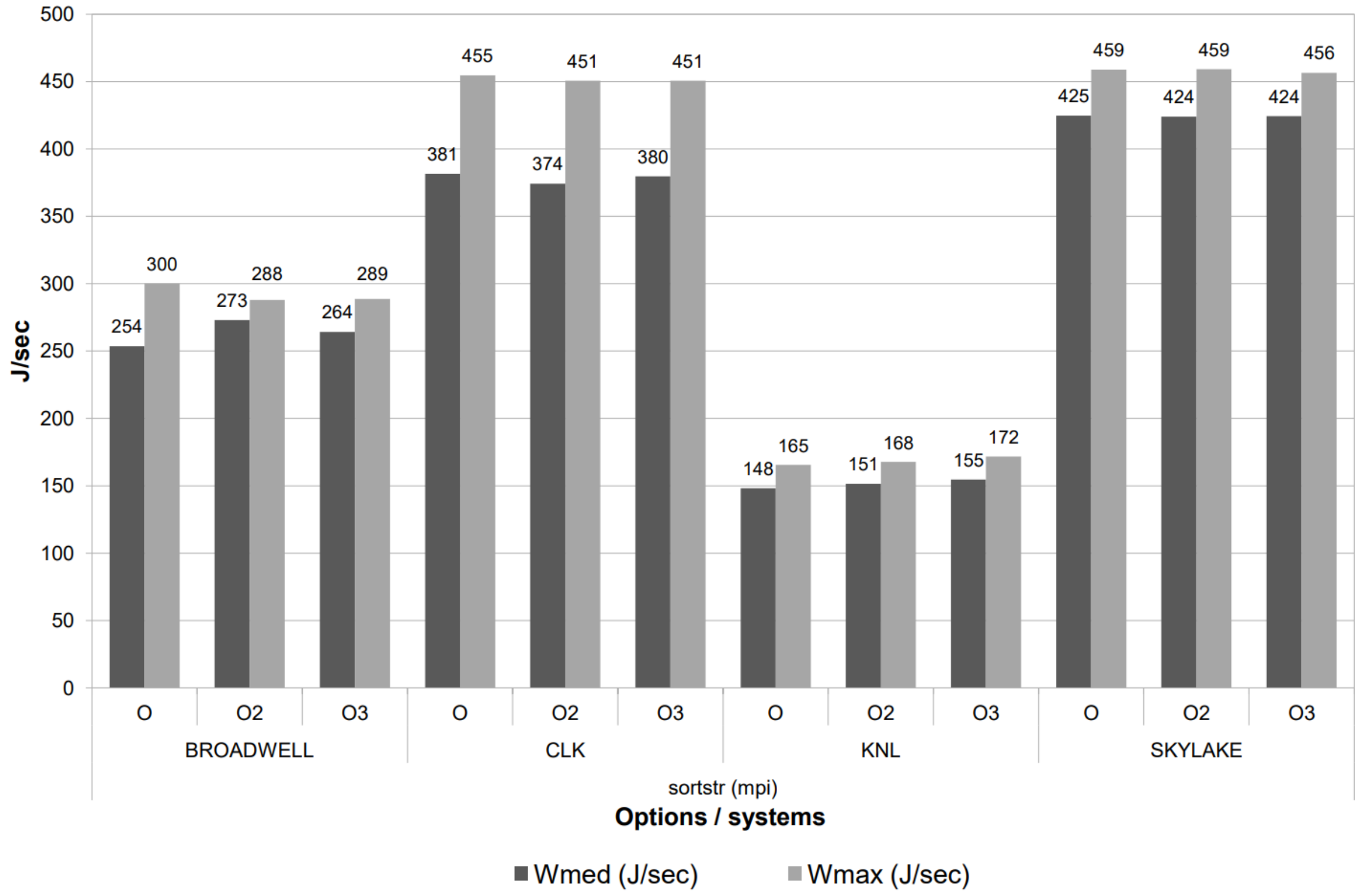

Figure 5. Peak and median power consumption depending on computer architecture and optimization flags for sortstr program

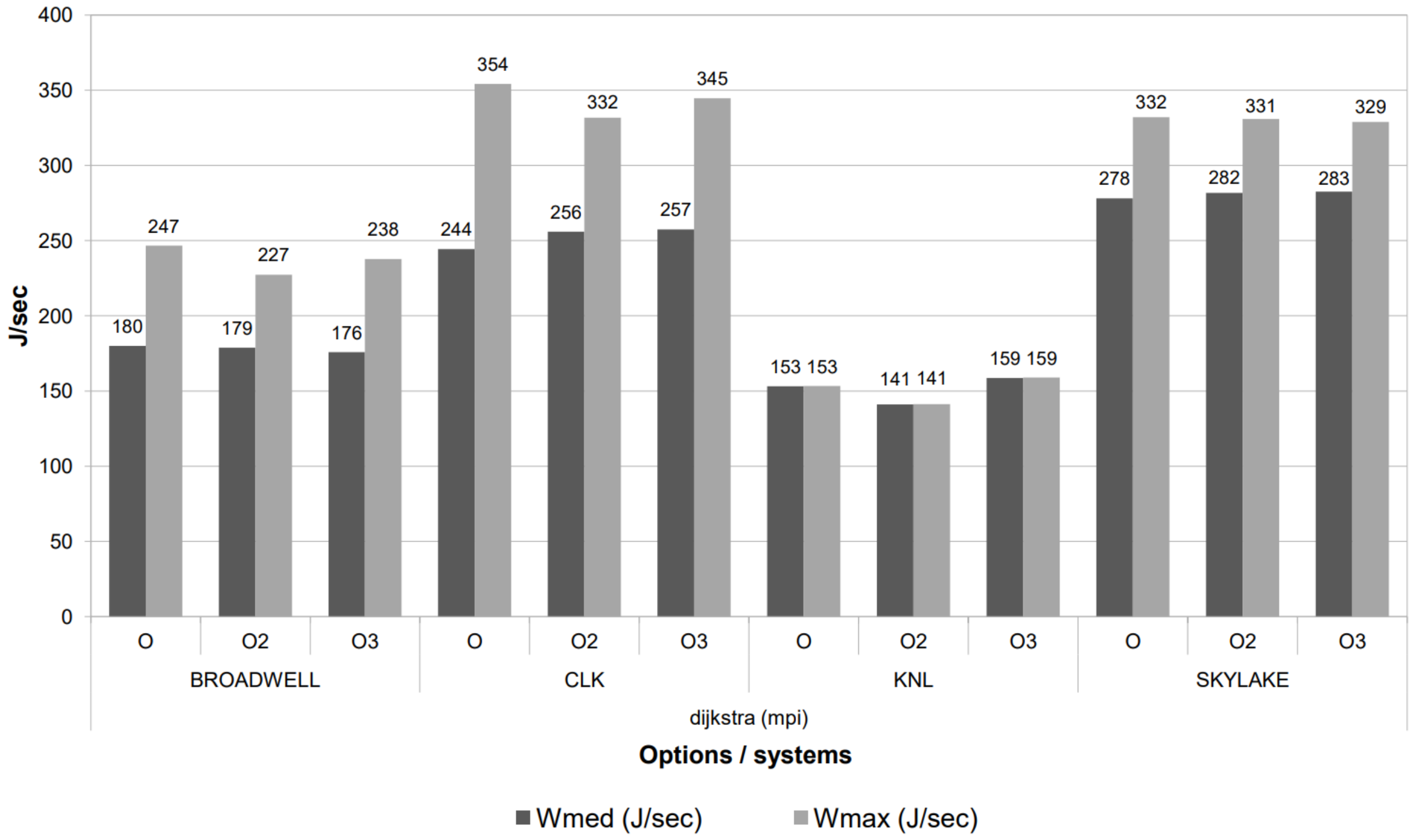

Figure 6. Peak and median power consumption depending on computer architecture and optimization flags for dijkstra program

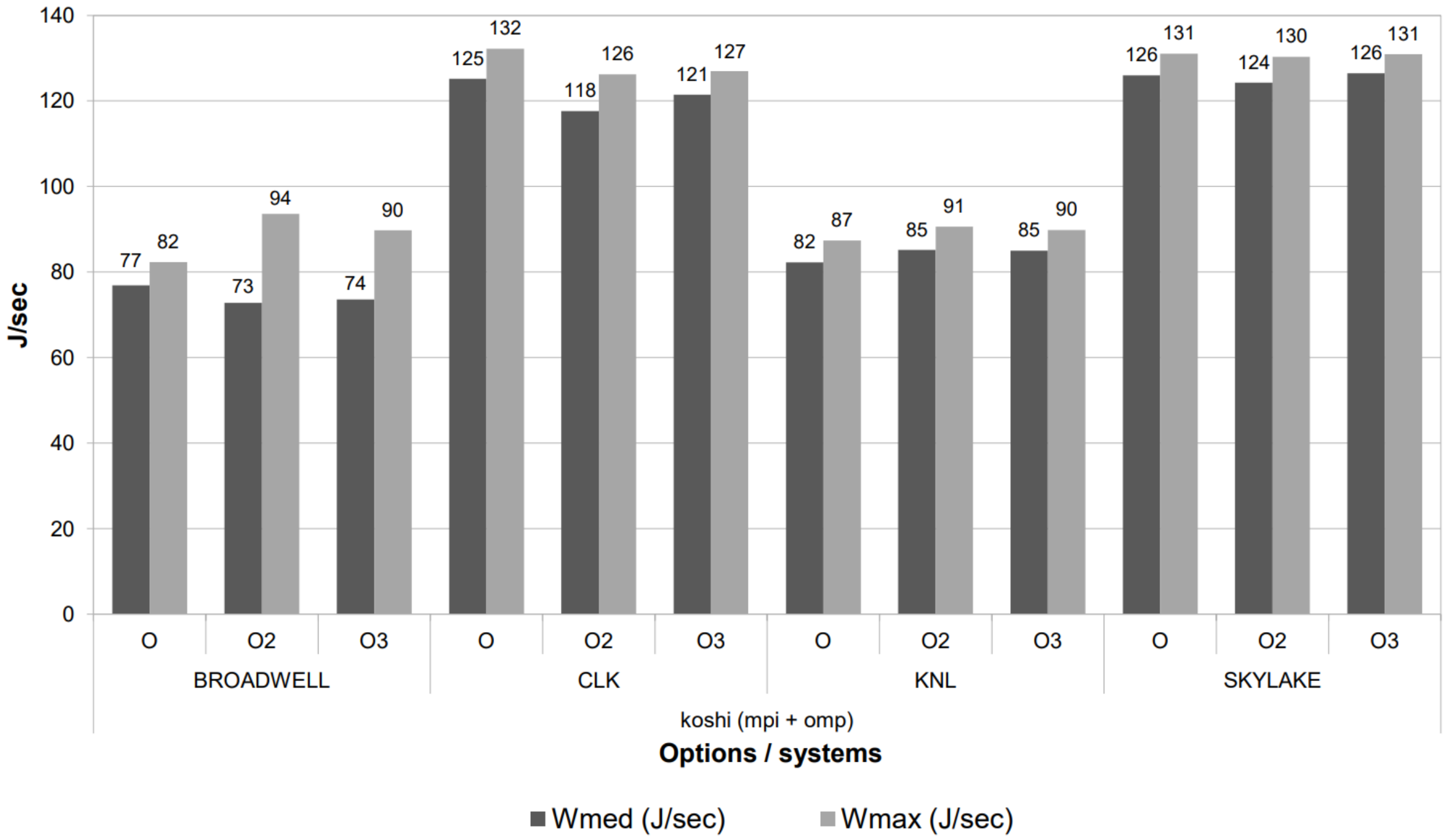

Figure 7. Peak and median power consumption depending on computer architecture and optimization flags for koshi program

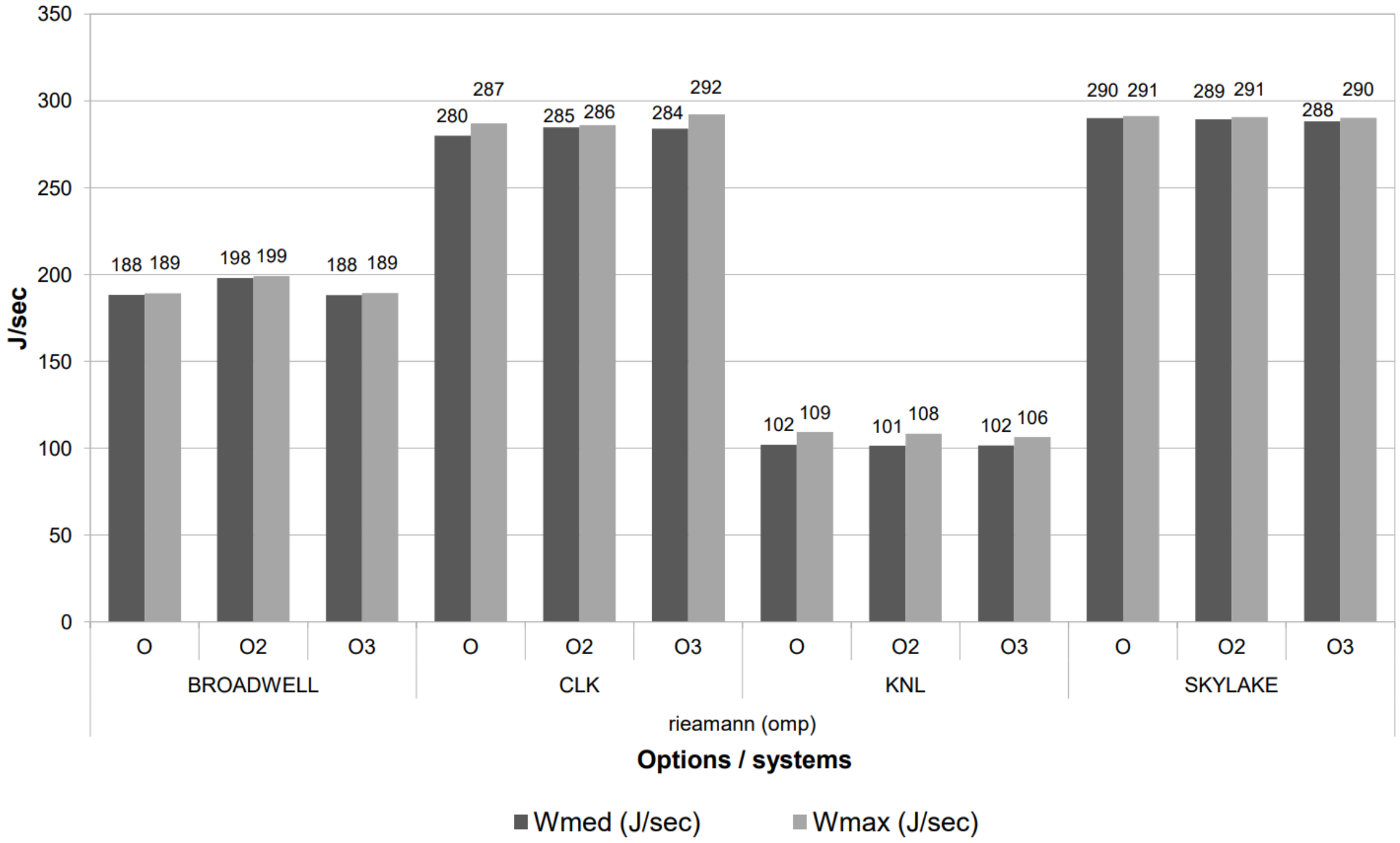

Figure 8. Peak and median power consumption depending on computer architecture and optimization flags for riemann program

## 6. CONCLUSIONS

Optimizing program code can significantly shorten program execution time and, as a result, reduce energy costs for its execution. However, the results of experiments show, than program optimization can result in an increase in the peak power consumption of the microprocessor. This occurs because at program optimization more complex and resource-intensive algorithms may be used, which consume more energy when performing computations. Memory operations also influence power consumption. Both analysis of previous studies and our results show that the highest peak power consumption values are achieved in parallel programs with intensive memory usage.

It was experimentally established that for programs with optimized source code, the average power consumption value tends to peak. Programs of that kind run faster and provide the greatest microprocessor load. This optimization has side effects such as an increase in microprocessor temperature. Comparing the median and peak power consumption values for application runs allows you to evaluate the efficiency of computing resources.

Analysis of power consumption graphs of a computer executing various parallel applications allows us to make an assumption about the possibility of creating program energy consumption profile that will be reproduced on different computing systems. Using this profile allows better planning the order of running parallel applications with high load and balancing the load on the computing system.

Acknowledgments. The work was carried out at the JSCC RAS as part of the government assignment (topic FNEF-2022-0016). Supercomputer MVS-10P OP was used.